\documentclass[iop,apj]{emulateapj}
\usepackage{natbib}
\usepackage{times}

\newcommand{\boltz}{{\tt BOLTZTRAN}}
\newcommand{\agile}{{\it Agile}}
\newcommand{\aboltz}{\agile-\boltz}

\newcommand{\msun}{\ensuremath{M_\sun}}

\newcommand{\nue}{\ensuremath{\nu_{e}}}
\newcommand{\nuebar}{\ensuremath{\bar \nu_e}}
\newcommand{\numt}{\ensuremath{\nu_{\mu\tau}}}
\newcommand{\numtbar}{\ensuremath{\bar \nu_{\mu\tau}}}
\newcommand{\numu}{\ensuremath{\nu_{\mu}}}
\newcommand{\nutau}{\ensuremath{\nu_{\tau}}}
\newcommand{\numubar}{\ensuremath{\bar \nu_{\mu}}}
\newcommand{\nutaubar}{\ensuremath{\bar \nu_{\tau}}}
\newcommand{\nux}{\ensuremath{\nu_x}}

\newcommand{\mev}{\mbox{MeV}}

\newcommand{\lepcons}{\ensuremath{N_{\rm L,cons}}}
\newcommand{\Ynu}{\ensuremath{Y_{\nu}}}
\newcommand{\Mshock}{\ensuremath{M_{\rm sh}}}
\newcommand{\meanE}[1]{\ensuremath{\langle E_{#1}\rangle_{\rm RMS}}}

\newcommand{\isotope}[2]{$^{#2}$#1}
\newcommand{\gcc}{\ensuremath{{\mbox{g~cm}}^{-3}}}

\newcommand{\Bethes}{\ensuremath{{\mbox{Bethe~s}}^{-1}}}

\newcommand{\momspder}{LHS/Term 2}
\newcommand{\fullop}{{{FullOp}}}
\newcommand{\reducop}{{ReducOp}}
\newcommand{\grfull}{GR-\fullop}
\newcommand{\nfull}{N-\fullop}
\newcommand{\nreduc}{N-\reducop}
\newcommand{\noc}{N-\reducop-NOC}

\newcommand{\chimera}{CHIMERA}
\newcommand{\vtd}{V2D}
\newcommand{\vertex}{V{\sc ertex}}
\newcommand{\vulcan}{{Vulcan/2D}}
\newcommand{\zidsa}{Zeus+IDSA}

\shorttitle{Neutrino Transport for Core-collapse Supernovae}
\shortauthors{Lentz et al.}

\begin{document}

\slugcomment{Accepted for publication in the Astrophys. J.}

\title{On the Requirements for Realistic Modeling of Neutrino Transport in
\\Simulations of Core-Collapse Supernovae}

\author{
Eric J. Lentz\altaffilmark{1,2,3},  
Anthony Mezzacappa\altaffilmark{2,1,4},\\
O.E. Bronson Messer\altaffilmark{5,1,4}, 
Matthias Liebend\"{o}rfer\altaffilmark{6}, 
W. Raphael Hix\altaffilmark{2,1}, 
and 
Stephen W. Bruenn\altaffilmark{7}
}
\email{elentz@utk.edu, mezzacappaa@ornl.gov}

\altaffiltext{1}{Department of Physics and Astronomy, University of Tennessee, Knoxville, TN 37996-1200, USA}
\altaffiltext{2}{Physics Division, Oak Ridge National Laboratory, P.O. Box 2008, Oak Ridge, TN 37831-6354, USA}
\altaffiltext{3}{Joint Institute for Heavy Ion Research, Oak Ridge National Laboratory, P.O. Box 2008, Oak Ridge, TN 37831-6374, USA}
\altaffiltext{4}{Computer Science and Mathematics Division, Oak Ridge National Laboratory, P.O.Box 2008, Oak Ridge, TN 37831-6164, USA}
\altaffiltext{5}{National Center for Computational Sciences, Oak Ridge National Laboratory, P.O. Box 2008, Oak Ridge, TN 37831-6164, USA}
\altaffiltext{6}{Department of Physics, University of Basel, Klingelbergstrasse 82, CH-4056 Basel, Switzerland}
\altaffiltext{7}{Department of Physics, Florida Atlantic University, 777 Glades Road, Boca Raton, FL 33431-0991, USA}

\begin{abstract}

We have conducted a series of numerical experiments with the spherically symmetric, general relativistic, neutrino radiation hydrodynamics code \aboltz\ to examine the effects of several approximations used in multidimensional core-collapse supernova simulations.
Our code permits us to examine the effects of these approximations quantitatively by removing, or substituting for, the pieces of supernova physics of interest.
These approximations include:
(1) using Newtonian versus general relativistic gravity, hydrodynamics, and transport;
(2) using  a reduced set of weak interactions, including the omission of non-isoenergetic neutrino scattering, versus the current state-of-the-art; and
(3) omitting  the velocity-dependent terms, or observer corrections,  from the neutrino Boltzmann kinetic equation.
We demonstrate that each of these changes has noticeable effects on the outcomes of our simulations.
Of these, we find that the omission of observer corrections is particularly detrimental to the potential for neutrino-driven explosions and exhibits a failure to conserve lepton number.
Finally, we discuss the impact of these results on our understanding of current, and the requirements for future, multidimensional models.

\end{abstract}

\keywords{methods: numerical --- neutrinos --- radiative transfer  --- supernovae: general}

\section{Introduction}

Colgate and White \citeyearpar{CoWh66}  were the first to propose that core-collapse supernovae may be neutrino-driven and performed the first numerical simulations of such events, launching more than four decades of research that continues to this day.
A significant milestone occurred nearly two decades later with Wilson's discovery that delayed neutrino-driven explosions could be obtained.
Based on his models, Wilson concluded \citep{Wils85,BeWi85} that the stalled supernova shock wave could be revived via neutrino absorption on a time scale of several hundred milliseconds given the intense flux of neutrinos emerging from the proto-neutron star liberating the star's gravitational binding energy.
Observations of the neutrinos from SN1987A, the first such observations of supernova neutrinos \citep{BiBlBr87,HiKaKo87}, provided support for the central role of neutrinos in the explosion mechanism.
State-of-the-art simulations today continue to explore Wilson's neutrino-driven explosion mechanism in the context of two- and three-dimensional models \citep[e.g., see][]{BuLiDe07,MaJa09,BrMeHi09b,SuKoTa10}.

Neutrinos are weakly interacting particles whose cross sections are energy dependent. Thus, unlike all other components in a supernova model, they are not well described as a fluid, except in the deepest layers, and their transition in space to non-fluid-like behavior  depends on their energy. Instead, the evolution of the neutrino radiation field, particularly in the semi-transparent regime, is far better characterized by classical kinetics---specifically, the general relativistic Boltzmann kinetic equation  \citep[e.g., see][]{CaMe03},
\notetoeditor{Superscripts should be placed to the left of subscripts in the following equation:}
\begin{equation}
p^{\hat \mu} \left(  {\Lambda^{\bar\mu}}_{\hat\mu} {e^\mu}_{\bar\mu} \frac{\partial f}{\partial x^\mu} - {\Gamma^{\hat\nu}}_{\hat\rho \hat\mu} p^{\hat\rho} \frac{\partial f}{\partial p^{\hat\nu}}\right) ={C}[f], \label{eq:boltz}
\end{equation}
where, for spherically symmetry,
\begin{eqnarray}
                \frac{1}{E}{C}[f]
                &=& \label{eq:collision}
                (1-f)j-\chi f \\ \nonumber
                &+&\frac{1}{c}\frac{1}{h^{3}c^{3}}E^{2}\int d\mu' R_{{\rm IS}}(\mu, \mu',E)f \\ \nonumber
                &-&\frac{1}{c}\frac{1}{h^{3}c^{3}}E^{2}f\int d\mu'   R_{{\rm IS}} (\mu, \mu',E)\\ \nonumber
                &+&\frac{1}{h^{3}c^{4}}
                       (1-f)  
                      \int dE'E'^{2}d\mu'  
                      R_{{\rm NIS}}^{{\rm in}}(\mu, \mu',E,E') f \\ \nonumber
                &-&\frac{1}{h^{3}c^{4}}
                      f  
                      \int dE'E'^{2}d\mu' 
                      R_{{\rm NIS}}^{{\rm out}}(\mu, \mu',E,E') (1-f) \\ \nonumber
                &+&\frac{1}{h^{3}c^{4}}
                       (1-f)  
                      \int dE'E'^{2}d\mu' 
                      R_{{\rm PR}}^{{\rm in}}(\mu, \mu',E,E') (1-\bar{f}) \\ \nonumber
                &-&\frac{1}{h^{3}c^{4}}
                      f  
                      \int dE'E'^{2}d\mu' 
                      R_{{\rm PR}}^{{\rm out}}(\mu, \mu',E,E') \bar{f}.
\end{eqnarray}
Equation (\ref{eq:boltz}) describes the evolution of the neutrino distribution function $f(t,x_1,x_2,x_3,\mu_1,\mu_2,E)$, which at time $t$ and spatial location $(x_1,x_2,x_3)$ supplies the distribution of neutrinos in direction cosines $(\mu_1,\mu_2)$ and energy $E$--i.e., the angular and spectral distribution of neutrinos. One such Boltzmann equation is solved for each flavor of neutrino---electron, muon, and tau (\nue, \numu, and \nutau, respectively)---and for their antineutrino partners (\nuebar, \numubar, and \nutaubar).
The invariant collision term, ${C}[f]$, in equation~(\ref{eq:collision}) is written using emission, $j$,  absorption, $\chi$, and scattering and pair kernels, $R$, following the forms often used for neutrino transport \citep[e.g., see][]{MeMe99}, where $\bar{f}$ is the distribution function for the partner antineutrino and $\mu$  is the neutrino direction cosine.
In equations (\ref{eq:boltz}) and (\ref{eq:collision}), $f$ is a function of $(\mu,E)$, as well as position and time.
The $(\mu',E')$ dependence of $f$ and $\bar{f}$ inside the integrals illustrates the physical coupling of all energies and angles for each neutrino species and of neutrino and antineutrino partners.

The first term on the left-hand side of equation (\ref{eq:boltz}) describes the time evolution of the local neutrino distribution owing to spatial transport through the volume of interest.
The second, far more complex term on the left-hand side (\momspder) describes the evolution of the local neutrino distribution in angle and energy as the result of (A) the coordinate system chosen, (B) special relativistic effects, and (C) general relativistic effects. In what follows, we will refer to (B) as ``observer corrections.''

Terms describing (A) depend on the choice of coordinate system. For example, in spherical-polar coordinates, the neutrino direction cosine relative to the outwardly pointing radial vector changes as the neutrino propagates through a local volume. This coordinate-system effect is included in \momspder\ and is present even in the absence of fluid motion or general relativity. For Cartesian coordinates, the neutrino direction cosines do not change as a result of the coordinate system choice alone and, consequently, such a term is absent.

Terms describing (B) depend on the frame of reference chosen to measure the neutrino direction cosines and energies.
The comoving frame, with neutrino direction cosines and energies  measured in an inertial frame of reference instantaneously comoving with the stellar core fluid with which the neutrinos interact, is often used.
Neutrino--matter interactions are naturally expressed in this frame.
Given this choice, the  terms in \momspder\  present a significant numerical challenge.
Finding discrete representations that guarantee conservation of lepton number and energy is one of the most difficult aspects of modeling neutrino transport in stellar cores. 
This has been achieved for general relativistic, spherically symmetric flows \citep{LiMeMe04}, providing the conceptual and implementation groundwork for achieving the same  in axisymmetric (2D) and non-symmetric (3D) flows. 
Further theoretical foundations have been laid \citep{CaMe03,CaLeMe05};
steps toward the development of a 2D Boltzmann solver have been taken \citep{OtBuDe08};
and the challenge now is to fully implement lepton energy and number conserving discretizations in 2D and 3D models.

For another choice of reference frame---measuring neutrino angles and energies relative to the inertial, ``lab" frame of a distant observer---the terms encapsulating the special relativistic effects in \momspder\ are absent, simplifying the left-hand side of the Boltzmann equation.
In such a frame of reference, the neutrino direction cosines and energies do not change from observer to observer in the frame.
However, this simplification comes at a price because the neutrino--matter interactions are naturally described in the comoving frame. 
In the lab frame,  a Lorentz transformation is required in order to express the comoving-frame neutrino--matter interactions in terms of the lab-frame direction cosines and energies, which introduces non-trivial velocity dependencies into the lab-frame collision term. 

One approach to the complexity of the lab-frame collision term is the ``mixed frame'' approach, which uses the lab-frame 4-momenta and an $\mathcal{O}(v/c)$ Taylor-series expansion in energy of the comoving-frame emissivities and opacities \citep{MiKl82}.
\citet{HuBu07} have proposed to use the mixed-frame approach for core-collapse simulations with extensions for non-isotropic and non-isoenergetic scattering.
The mixed-frame approach may be difficult to extend to arbitrarily relativistic flows, and has not yet been used in the context of a full-physics core-collapse supernova simulation.

In a general relativistic setting, such as core-collapse supernovae, we must contend with (B) and/or (C) regardless of the frame of reference chosen to describe the neutrino direction cosines and energies.
Even for static general relativistic environments, angular aberration, gravitational red shift, and other effects occur, and the resulting terms in (C) are always present. 

Regardless of  approach, comoving- or lab-frame, it is problematic to adapt the simplicity of both approaches, simultaneously  simplifying the left- and right-hand sides of the Boltzmann equation, as has been done in \citet{BuLiDe06,BuLiDe07}, \citet{OtBuDe08}, and other models using the \vulcan\ code, although one can view the implementation in these works as steps toward a more complete description.
They deploy a lab-frame approach for terms describing angular aberration and energy shift on the left-hand side (or assume such terms are unimportant in a comoving-frame approach), while simultaneously deploying a comoving-frame approach for the collision term describing the neutrino--matter interactions on the right-hand side. This is not a mixed-frame approach in the sense described above.
It is an approach not based in any reference frame, and it is physical only for static cases in which there is no distinction between lab and comoving frames.
One of the goals of this study is to investigate the importance of the terms in \momspder\ in a comoving-frame approach, and whether they can be ignored while using a comoving-frame approach for the collision term.

Modeling general relativistic Boltzmann kinetics is also challenging because of the complexity of the collision term on the right-hand side of the Boltzmann equation, even in a comoving-frame formulation. Looking at equation (\ref{eq:collision}), we see that the collision term describes the full, direct coupling of all neutrino angles and energies for each neutrino species, owing to neutrino isoenergetic (IS) scattering on nuclei and non-isoenergetic (NIS) scattering on electrons and nucleons.
The pair creation and annihilation processes (PR) such as electron--positron annihilation and nucleon--nucleon bremsstrahlung also couple the angles and energies of the neutrino and antineutrino species of  each  flavor together.
The coupling of all neutrino angles and energies through the relevant set of weak interactions dominates the computation associated with the solution of the neutrino Boltzmann equations.
It has been argued \citep{BuLiDe06,BuLiDe07,NoBuAl10} these couplings are subdominant and can be ignored,  greatly simplifying the neutrino Boltzmann equations and significantly reducing the computational cost associated with their solution. A second goal of this study is to investigate whether or not such  approximations to the collision term are realistic.

The complete general relativistic Boltzmann equation was solved in spherically symmetric models of core-collapse supernovae by the Oak Ridge-Basel collaboration \citep{LiMeTh01,LiMeMe04} and  by Sumiyoshi and collaborators \citep{SuYaSu05}.
Achieving this  in three-dimensional models of core-collapse supernovae presents a major challenge, one that will likely require sustained exascale resources to meet.

The overarching goal of this study is to use  general relativistic, spherically symmetric Boltzmann simulations to guide and, more importantly, set minimum requirements for accurate 2D and 3D simulations.
We use the Oak Ridge-Basel code \aboltz\  in these studies to compare general relativistic--full weak interaction physics (\grfull), Newtonian--full weak interaction physics (\nfull), Newtonian--reduced weak interaction physics (\nreduc), and Newtonian--reduced weak interaction physics--no observer correction (\noc) models.
These models will demonstrate the importance of general relativity, a complete weak interaction set and treatment, and the terms in \momspder\ to stellar core collapse and the post-core-bounce evolution. Current multidimensional models suggest that spherical symmetry is a reasonable approximation for the first 100--150 ms after bounce \citep{MaJa09,BrMeHi09b,SuKoTa10}. Thus, the simulations presented here are relevant for discussing the initial conditions present for all multidimensional phenomena that might ensue; e.g., neutrino-driven convection and the standing accretion shock instability (SASI).

\section{Disabling Observer Corrections in a Lagrangian Formulation}\label{sec:noc}

Disabling general relativity in a simulation, instead running a Newtonian simulation, is straightforward and requires no special considerations to define or interpret. The same holds true for limiting the weak interaction channels included in the collision term on the right hand side of the Boltzmann equation. Newtonian or general relativistic simulations can be performed with more, or less, weak interaction physics.
However, disabling the observer corrections in a model requires some definition and care. 

Using \citet{MeMa89}, equation (VI.11), we begin by expressing the neutrino Boltzmann equation in flat spacetime, Eulerian spherical-polar spacetime coordinates with zero shift vector, comoving-frame 4-momenta, and in nonconservative form:
\notetoeditor{Powers should be placed to the right of `0' subscripts in the equations for the rest of this section: ${E_0}^3$, ${\mu_0}^2$}
\begin{eqnarray}
\lefteqn{\frac{\partial f}{\partial \tilde{t}} \label{eq:boltzeul}
+ \frac{\mu_0 + v}{1+\mu_0 v}\frac{\partial f}{\partial r} 
}\\ \nonumber
 &+&  \left[ \frac{1}{r}-\gamma^2 \left( \frac{\partial v}{\partial \tilde{t}}+ \frac{\mu_0 + v}{1+\mu_0 v}\frac{\partial v}{\partial r}\right)\right] \left(1-{\mu_0}^2\right)\frac{\partial f}{\partial \mu_0} \\ \nonumber
&-& \left[\frac{1-{\mu_0}^2}{1+\mu_0 v}\frac{v}{r} 
+ \mu_0 \gamma^2\left( \frac{\partial v}{\partial \tilde{t}}+ \frac{\mu_0 + v}{1+\mu_0 v} \frac{\partial v}{\partial r}\right)\right]E_0\frac{\partial f}{\partial E_0}\\ \nonumber
&  &{}=\frac{1}{\gamma E_0}\frac{1}{1+\mu_0 v}(e-of)\\ \nonumber
&  &{}\equiv \frac{1}{\gamma E_0}\frac{1}{1+\mu_0 v} C[f]. \nonumber
\end{eqnarray}
In equation (\ref{eq:boltzeul}), $\mu_0$ and $E_0$ are the neutrino direction cosine and energy as measured in a comoving frame of reference, and $e$ and $o$ are  the invariant emissivity and opacity. 
We use $c=1$ throughout this section, and have written the Eulerian time coordinate as $\tilde{t}$.
Multiplying by $(1+\mu_0 v)$ and rearranging we have
\begin{eqnarray}
& & (1+\mu_0 v)\frac{\partial f}{\partial \tilde{t}} \label{eq:boltzeulov/c}
+ (\mu_0 + v)\frac{\partial f}{\partial r} 
+ \frac{1-{\mu_0}^2}{r}\frac{\partial f}{\partial \mu_0}\\ \nonumber
& + & \left[\left(\frac{v}{r} - \frac{\partial v}{\partial r}\right)\mu_0 \left(1-{\mu_0}^2\right)+\mathcal{O}\left(v^2\right)\right]\frac{\partial f}{\partial \mu_0} \\ \nonumber
&+& \left[{\mu_0}^2 \left(\frac{v}{r}-\frac{\partial v}{\partial r}\right)-\frac{v}{r}+\mathcal{O}\left(v^2\right)\right]E_0\frac{\partial f}{\partial E_0} = \frac{1}{E_0}C[f]. \nonumber
\end{eqnarray}
Using the continuity equation,
\begin{displaymath}
\frac{1}{\rho}\frac{D\rho}{D\tilde{t}}+\frac{3v}{r}=\frac{v}{r}-\frac{\partial v}{\partial r},
\end{displaymath}
where $D/D\tilde{t}=\partial /\partial \tilde{t} + v\partial /\partial r$,
equation (\ref{eq:boltzeulov/c}), and dropping $\mathcal{O}(v^2)$ terms we get
\begin{eqnarray}
\lefteqn{(1+\mu_0 v)\frac{\partial f}{\partial \tilde{t}} \label{eq:boltzeulcont}
+ (\mu_0 + v)\frac{\partial f}{\partial r} 
+ \frac{1-{\mu_0}^2}{r}\frac{\partial f}{\partial \mu_0}}\\ \nonumber
&  &{}+ \left(\frac{1}{\rho}\frac{D\rho}{D\tilde{t}}+\frac{3v}{r}\right)\mu_0 \left(1-{\mu_0}^2\right)\frac{\partial f}{\partial \mu_0} \\ \nonumber
&& {}+ \left[{\mu_0}^2 \left(\frac{1}{\rho}\frac{D\rho}{D\tilde{t}}+\frac{3v}{r}\right)-\frac{v}{r}\right]E_0\frac{\partial f}{\partial E_0}=\frac{1}{E_0}C[f]. \nonumber
\end{eqnarray}
We can express the observer correction terms in conservative form using
\begin{displaymath}
\mu_0 \left(1-{\mu_0}^2\right)\frac{\partial f}{\partial \mu_0}=\frac{\partial\left[\mu_0 \left(1-{\mu_0}^2\right)f\right]}{\partial \mu_0}-f \left(1-3{\mu_0}^2\right)
\end{displaymath}
and
\begin{displaymath}
E_0\frac{\partial f}{\partial E_0}=\frac{1}{{E_0}^2}\frac{\partial \left({E_0}^3f\right)}{\partial E_0}-3f.
\end{displaymath}
Substituting these expressions into equation (\ref{eq:boltzeulcont}), we get the $\mathcal{O}(v)$  Boltzmann equation,
\begin{eqnarray}
\lefteqn{(1+\mu_0 v)\frac{\partial f}{\partial \tilde{t}} \label{eq:boltzeulov/ccon}
+ (\mu_0 + v)\frac{\partial f}{\partial r} 
+ \frac{1-{\mu_0}^2}{r}\frac{\partial f}{\partial \mu_0} }\\ \nonumber
& + &\left(\frac{1}{\rho}\frac{D\rho}{D\tilde{t}}+\frac{3v}{r}\right)\frac{\partial \left[\mu_0 \left(1-{\mu_0}^2\right)f\right]}{\partial \mu_0}\\ \nonumber
&+& \left[{\mu_0}^2 \left(\frac{1}{\rho}\frac{D\rho}{D\tilde{t}}+\frac{3v}{r}\right)-\frac{v}{r}\right]\frac{1}{{E_0}^2}\frac{\partial \left({E_0}^3f\right)}{\partial E_0}
-\frac{1}{\rho}\frac{D\rho}{D\tilde{t}}f\\ \nonumber
&&{}=\frac{1}{E_0}C[f].
\end{eqnarray}
Note, the last term on the left hand side of equation (\ref{eq:boltzeulov/ccon}), $-(f/\rho) D\rho/D\tilde{t}$, is part of the observer corrections. 

Our Eulerian starting point with comoving-frame neutrino 4-momenta (equation \ref{eq:boltzeulov/c}) provides the formulation in which we can most readily discuss what it means to have Boltzmann neutrino transport without observer corrections.
We define the evolution of the neutrino distributions in this case to be governed by
\begin{equation}
(1+\mu_0 v)\frac{\partial f}{\partial \tilde{t}} \label{eq:boltzeulnoc}
+ (\mu_0 + v)\frac{\partial f}{\partial r} 
+ \frac{1-{\mu_0}^2}{r}\frac{\partial f}{\partial \mu_0} = \frac{1}{E_0}C[f];
\end{equation}
that is, by equation (\ref{eq:boltzeulov/c}) with the velocity-dependent, neutrino angle- and energy-shift terms ignored.

From equations (15)--(22) of \citet{LiMeMe04}, \aboltz\ evolves the following purely Lagrangian (comoving-frame spacetime coordinates and neutrino 4-momenta) equations in flat spacetime,
\begin{eqnarray}
\lefteqn{\frac{\partial F}{\partial t} \label{eq:boltzlag}
+ 4\pi\mu_0\frac{\partial \left(r^2\rho F\right)}{\partial m} 
+ \frac{1}{r}\frac{\partial \left[\left(1-{\mu_0}^2\right)F\right]}{\partial \mu_0} } \\ \nonumber
& &{}+ \left(\frac{1}{\rho}\frac{\partial\rho}{\partial t}+\frac{3v}{r}\right)\frac{\partial \left[\mu_0 \left(1-{\mu_0}^2\right)F\right]}{\partial \mu_0} \\ \nonumber
&&{}+ \left[{\mu_0}^2 \left(\frac{1}{\rho}\frac{\partial\rho}{\partial t}+\frac{3v}{r}\right)-\frac{v}{r}\right]\frac{1}{{E_0}^2}\frac{\partial \left({E_0}^3F\right)}{\partial E_0} = \frac{1}{E_0}C[F], 
\end{eqnarray}
where $F=f/\rho$ is the specific neutrino distribution function.
Note, here the Lagrangian partial derivatives with respect to $t$ and $m$ are at constant $m$ and $t$, respectively.
We can express the partial derivatives with respect to the Lagrangian spacetime coordinates, $(t,m)$, in terms of the Eulerian spacetime coordinates, $(\tilde{t},r)$, using
\begin{displaymath}
\frac{\partial }{\partial t} = \frac{\partial \tilde{t}}{\partial t}\frac{\partial }{\partial \tilde{t}}
                                          +\frac{\partial r}{\partial t}\frac{\partial }{\partial r}
\end{displaymath}
and
\begin{displaymath}
\frac{\partial }{\partial m} = \frac{\partial \tilde{t}}{\partial m}\frac{\partial }{\partial \tilde{t}}
                                          +\frac{\partial r}{\partial m}\frac{\partial }{\partial r}.
\end{displaymath}
Using equation (45) of \citet{CaMe03}, we have to $\mathcal{O}(v)$,
\begin{displaymath}
\frac{\partial }{\partial t} = \frac{\partial }{\partial \tilde{t}}
                                          +v\frac{\partial }{\partial r} = \frac{D}{D\tilde{t}}
\end{displaymath}
and
\begin{displaymath}
\frac{\partial }{\partial m} = \frac{v}{4\pi r^2\rho}\frac{\partial }{\partial \tilde{t}}
                                          +\frac{1}{4\pi r^2\rho}\frac{\partial }{\partial r}.
\end{displaymath}
Substituting these  transformations into the first three terms in equation (\ref{eq:boltzlag}) gives
\begin{eqnarray}
\lefteqn{\frac{\partial F}{\partial t} \label{eq:temp1}
+ 4\pi\mu_0\frac{\partial \left(r^2\rho F\right)}{\partial m}
+\frac{1}{r}\frac{\partial \left[\left(1-{\mu_0}^2\right)F\right]}{\partial \mu_0} } \\ \nonumber
& = & \frac{\partial F}{\partial \tilde{t}}+v\frac{\partial F}{\partial r}+\frac{\mu_0 v}{r^2 \rho}\frac{\partial \left(r^2\rho F\right)}{\partial \tilde{t}}
+\frac{\mu_0}{r^2 \rho}\frac{\partial \left(r^2\rho F\right)}{\partial r} \\ \nonumber
&&{}+\frac{1}{r}\frac{\partial \left[\left(1-{\mu_0}^2\right)F\right]}{\partial \mu_0}.\\ \nonumber
\end{eqnarray}
Writing the right hand side of equation (\ref{eq:temp1}) in terms of $f$ and expanding, we are left with
\begin{eqnarray}
\lefteqn{\frac{\partial F}{\partial t}
+ 4\pi\mu_0\frac{\partial (r^2\rho F)}{\partial m}
+\frac{1}{r}\frac{\partial \left[\left(1-{\mu_0}^2\right)F\right]}{\partial \mu_0} }\\ \nonumber
&=&
\frac{1}{\rho}\left[\left(1+\mu_0 v\right)\frac{\partial f}{\partial \tilde{t}} 
+ (\mu_0 + v)\frac{\partial f}{\partial r} 
+ \frac{1-{\mu_0}^2}{r}\frac{\partial f}{\partial \mu_0}\right]
 \nonumber \\
&& {}-\frac{f}{\rho^2}\frac{D\rho}{D \tilde{t}} + \mathcal{O}\left(v^2\right). \nonumber
\end{eqnarray}
Moving the term containing the Lagrangian time-derivative, $D /D \tilde{t}$,   of the density to the LHS and restating it using the Lagrangian time-derivative, $\partial /\partial t$, we have to $\mathcal{O}(v)$,
 \begin{eqnarray}
\lefteqn{ \frac{\partial F}{\partial t} 
+ 4\pi\mu_0\frac{\partial (r^2\rho F)}{\partial m}
+\frac{1}{r}\frac{\partial \left[\left(1-{\mu_0}^2\right)F\right]}{\partial \mu_0}
+\frac{F}{\rho}\frac{\partial \rho}{\partial t}  }\\ \nonumber
&&=
\frac{1}{\rho}\left[\left(1+\mu_0 v\right)\frac{\partial f}{\partial \tilde{t}} 
+ (\mu_0 + v)\frac{\partial f}{\partial r} 
+ \frac{1-{\mu_0}^2}{r}\frac{\partial f}{\partial \mu_0}\right].
\end{eqnarray}
Therefore, with equation (\ref{eq:boltzeulnoc}) as a guide, a ``no-observer-correction'' run in our Lagrangian formulation would correspond to a solution of the following equation:
\begin{eqnarray}
\frac{\partial F}{\partial t} \label{eq:lagnoc}
&+& 4\pi\mu_0\frac{\partial \left(r^2\rho F\right)}{\partial m}
+\frac{1}{r}\frac{\partial \left[\left(1-{\mu_0}^2\right)F\right]}{\partial \mu_0}
+\frac{F}{\rho}\frac{\partial \rho}{\partial t} \\
&&{}= \frac{1}{E_0}\frac{1}{\rho}C[f]\equiv \frac{1}{E_0}C[F]. \nonumber
\end{eqnarray}
Equation~(\ref{eq:lagnoc}) is used in our `no-observer-corrections' model, \noc.
Equation~(\ref{eq:lagnoc}) is clearly not manifestly conservative for neutrino number when integrated over mass and when the density evolves, and therefore its Eulerian equivalent, equation~(\ref{eq:boltzeulnoc}), is also not number conservative when integrated over volume.
On the other hand, equation (\ref{eq:boltzeulov/ccon}) {\it is} manifestly number conservative when integrated over volume (after dropping the $\mu_{0}v\partial f/\partial t$ term). The culprit in equation (\ref{eq:boltzeulnoc}) is the $v\partial f/\partial r$ term. When expressed in volume-conservative form, this term contains a velocity divergence, or equivalently a logarithmic time derivative of the density, that would normally be cancelled by the logarithmic time derivative of the density in equation (\ref{eq:boltzeulov/ccon}). But when the observer corrections are dropped, the last term on the LHS of equation (\ref{eq:boltzeulov/ccon}) is dropped, and this cancellation no longer occurs and we are left with the same term that appears as the last term on the LHS of equation (\ref{eq:lagnoc}), which breaks number conservation. By expressing our observer corrections in equation (\ref{eq:boltzeulov/ccon}) in conservative form, we made explicit this logarithmic time derivative of density contained within them.

\section{Numerical methods and inputs}

All models in this paper are computed using the parallel version of the general relativistic, spherically symmetric, neutrino radiation hydrodynamics code \aboltz\ \citep{LiMeMe04} with extensions that we describe here.

\begin{deluxetable*}{ccc}
\tabletypesize{\scriptsize}
\tablecaption{Neutrino Opacity Summary Table\label{tab:opac}}
\tablecolumns{3}
\tablewidth{0pt}
\tablehead{
\colhead{Interaction} & \colhead{\fullop\ Opacities } & \colhead{\reducop\ Opacities} 
}
\startdata
$\nu e^- \leftrightarrow \nu' e^-$   & \citet{SchSh82} & None  \\
$\nu e^+ \leftrightarrow \nu' e^+$ & &   \\
$\nu n \leftrightarrow \nu' n$          & \citet{RePrLa98} & \citet{Brue85}   \\
$\nu p \leftrightarrow \nu' p$          &&  \\
$e^- p \leftrightarrow \nu_e n $     &\citet{RePrLa98} & \citet{Brue85}   \\
 $e^+ n \leftrightarrow \bar{\nu}_e p $ &&   \\
 $\nu A \leftrightarrow \nu A$              & \citet{Brue85} & \citet{Brue85}  \\
$\nu \alpha \leftrightarrow \nu \alpha$ & \citet{Brue85} & \citet{Brue85}  \\
$e^- (A,Z) \leftrightarrow \nu_e (A,Z-1) $  &\citet{LaMa00} & \citet{Brue85}  \\
                                                                   &\citet{LaMaSa03} &    \\
$ e^-e^+ \leftrightarrow \nu \bar{\nu} $ & \citet{SchSh82}  & \citet{SchSh82}   \\
$ NN \leftrightarrow NN\nu\bar{\nu}$&  \citet{HaRa98}   & \citet{HaRa98} 
\enddata
\end{deluxetable*}

\subsection{\aboltz}

\aboltz\ is a combination of the general relativistic (GR) hydrodynamics code \agile\ \citep{LiRoTh02} and the neutrino transport code \boltz\ \citep{MeBr93b,MeMe99,LiMeMe04}.
\agile\ solves the complete GR spacetime and hydrodynamics equations implicitly in spherical symmetry on a dynamic, moving grid.
The moving grid allows adequate resolution of the shock using only $\mathcal{O}(100)$ radial zones.
Recent enhancements include the use of a TVD (total variation diminishing) hydrodynamics solver \citep{LiRaJa05}, which improves the accuracy of advection, and the use of  $\delta m$ as the grid coordinate rather than the enclosed mass  \citep[][ \S2.1]{FiWhMe10}, which improves numerical accuracy when mass zones are small and  density gradients are large.
In Newtonian mode the gravitational mass is set equal to the baryonic mass (omitting the non-rest-mass energy contributions) and the relativistic parameters are set to their non-relativistic values: $\alpha = 1, \Gamma =1$.
\boltz\ \citep{MeBr93b,MeMe99,LiMeMe04} solves the GR extension of the spectral neutrino Boltzmann equation (Eq.~\ref{eq:boltzlag}) with a  Gauss-Legendre ($S_N$) quadrature.
Here we use an 8-point angular quadrature  and 20 logarithmically-spaced energy groups with group centers from 3 to 300~\mev.
Previous studies \citep{MeBr93a,LiMeMe04} with  (\agile-)\boltz\ have shown that 20-group energy resolution is adequate in removing artifacts seen at lower (12-group) energy resolution, and their 12- and 20-group runs exhibited no differences in outcomes.
Moreover, 20-group energy resolution matches, or exceeds, the resolution used for supernova models computed with the multidimensional codes we discuss in \S\ref{sec:codes}.
The discretization scheme is designed to simultaneously conserve lepton number and energy as  described in \citet{LiMeMe04}.
Since we do not include any physics to distinguish between muon- and tau-flavored leptons, we use the combined species $\numt = \{\numu,\nutau\}$ and $\numtbar = \{\numubar,\nutaubar\}$.

For all models we use the nuclear, electron, and photon equations of state (EoS) of \citet{LaSw91} with the bulk incompressibility of nuclear matter $\kappa_s = 220 \, \mev$.\footnote{We use the latest version of the \citet{LaSw91} EoS, version 2.7, which is available for download from its authors at \url{http://www.astro.sunysb.edu/dswesty/lseos.html}.}
This matches the current experimental value of $\kappa_s = 240 \pm 20 \, \mev$ \citep{ShKoCo06} better than the  value of 180~\mev\ more commonly used with LS~EoS in the past, though the value of $\kappa_s$ in LS~EoS has been shown to be of little consequence during the early phases of core-collapse supernova evolution shown here \citep{SwLaMy94,ThBuPi03,LeHiBa10}.
Matter outside the ``iron'' core is treated as an ideal gas of \isotope{Si}{28} that  ``flashes'' instantaneously to nuclear statistical equilibrium when the temperature exceeds 0.47~\mev.

The stellar progenitor used for all models reported here is the 15-\msun\ solar-metalicity progenitor of \citet{WoHe07}. We have mapped the inner $1.8 \, \msun$ of the progenitor onto 108 mass shells of the adaptive radial grid.

\subsection{Neutrino Opacities}

The base, or full,  opacity set (\fullop) includes emission, absorption, and scattering on free nucleons \citep{RePrLa98}; isoenergetic scattering on $\alpha$-particles and heavy nuclei \citep{Brue85}; scattering of neutrinos on electrons  (NES) and positrons (NPS)  \citep{SchSh82}; production of neutrino pairs from $e^+e^-$ annihilation \citep{SchSh82} and nucleon-nucleon bremsstrahlung \citep{HaRa98}; and  electron capture (EC) on nuclei using the LMSH EC table of \citet{LaMaSa03}, which utilizes the EC rates of \citet{LaMa00}.
The full angle and  energy exchange for scattering between the neutrinos and  electrons, positrons, and nucleons is included, while  scattering on nuclei is isoenergetic (IS).
Bremsstrahlung and $e^-e^+$ annihilation are the only sources of \numt\ and \numtbar.

For our reduced opacity set (\reducop) we replace the LMSH EC table for electron capture on nuclei with  an independent particle approximation (IPA)  \citep{Full82} using the implementation described in \citet{Brue85}, which cuts off when the mean neutron number of the heavy nuclei  $N \geq 40$.
We also drop all scatterings (NIS) that couple neutrino-energy groups.
The primary contribution of electron and positron scattering opacities is through neutrino-energy down-scattering \citep{MeBr93c}, not through their contribution to the total scattering opacity; therefore,  we omit them completely from the \reducop\ opacity set.
We also replace the NIS nucleon scattering opacities of \citet{RePrLa98} with the more approximate IS equivalents from \citet{Brue85}.
For consistency, we also replace the neutrino emission and absorption opacities of \citet{RePrLa98} with their \citet{Brue85} equivalents.
Ion-ion correlations and weak magnetism are omitted from both opacity sets.
The two opacity sets are summarized in Table~\ref{tab:opac}.

\subsection{Observer Corrections}

As noted in \S\ref{sec:noc}, the Lagrangian formulation in \aboltz\ and the use of the specific neutrino distribution function, $F=f/\rho$, which is needed to properly account for number and energy conservation \citep[see discussion on the necessity of using $F$ for Lagrangian models in][ \S IV.B]{CaMe03}, require care in the definition of a no-observer-corrections model.
Moreover, time derivatives at  fixed Lagrangian mass coordinates on a moving grid must be handled with care \citep[see][ \S 3.2]{LiMeMe04}.
Therefore, for model \noc, we implement the ``compression'' term in the no-observer-corrections transport equation~(\ref{eq:lagnoc}) by re-expressing the time derivative of density as a spatial divergence, using the continuity equation,
\begin{equation}
\frac{F}{\rho} \frac{\partial \rho}{\partial t} = -\frac{F}{r^2} \frac{\partial \left(r^2 v\right)}{\partial r}.
\end{equation}

\section{Results}

\begin{figure*}
\epsscale{1}
\plotone{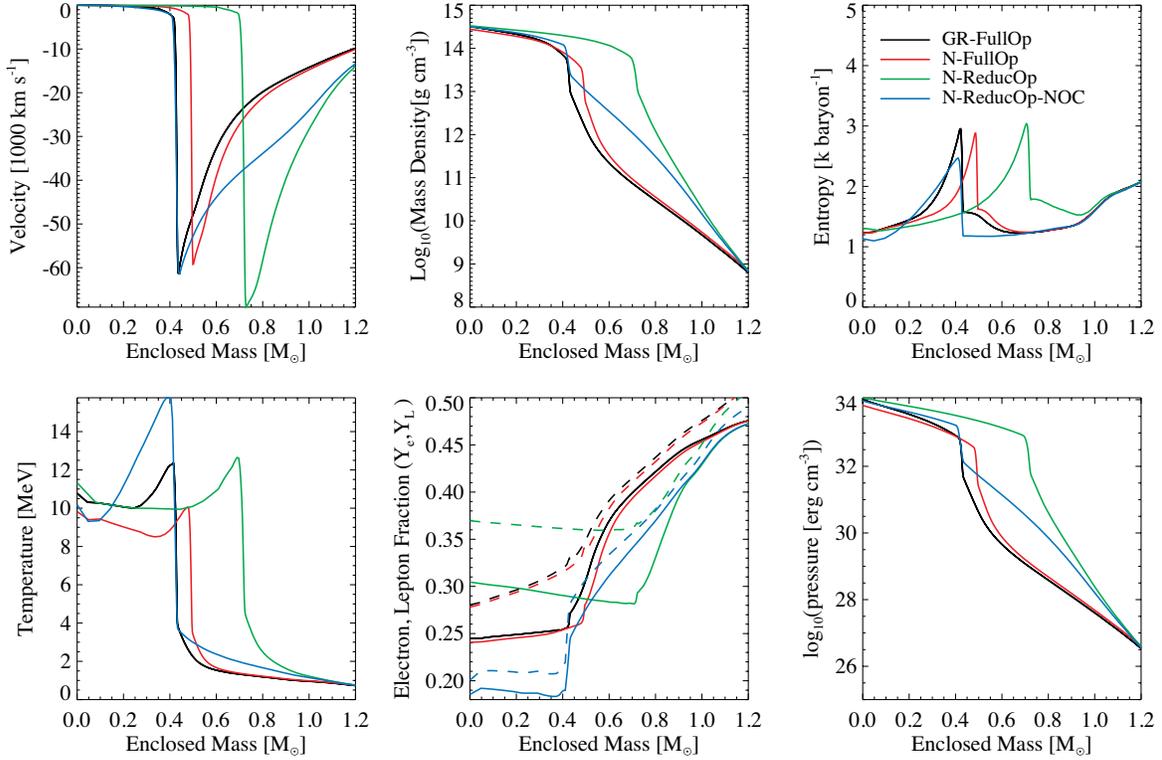}
\caption{Properties of our models at core bounce, where bounce is defined as the maximum compression of the central density during the launch of the bounce shock.
The models are: general relativistic gravity, hydrodynamics and transport  with full opacities (\grfull, plotted in black); Newtonian gravity with full opacities and $\mathcal{O}(v/c)$ hydrodynamics and transport (\nfull, plotted in red); Newtonian gravity with reduced opacities  and $\mathcal{O}(v/c)$ hydrodynamics and transport (\nreduc, plotted in green); and Newtonian gravity with $\mathcal{O}(v/c)$ hydrodynamics and reduced opacities, and $\mathcal{O}(1)$ transport (\noc, plotted in blue).
The panels are: radial velocity (upper left), density (upper center), entropy (upper right), temperature ($kT$, lower left), net electron (or proton) fraction ($Y_e$, lower center, solid lines), net lepton fraction ($Y_L = Y_e +  (n_{\nue} - n_{\nuebar})/n_{\rm baryons}$, lower center, dashed lines), and pressure (lower right). All quantities are plotted relative to enclosed rest-mass in \msun.   \label{fig:bounce}
}
\end{figure*}

\begin{deluxetable*}{ccccc}
\tabletypesize{\scriptsize}
\tablecaption{Model Approximations and Properties\label{tab:models}}
\tablecolumns{5}
\tablewidth{0pt}
\tablehead{
\colhead{Property} & \colhead{\grfull } & \colhead{\nfull} & \colhead{\nreduc} & \colhead{\noc}
}
\startdata
Gravity  and hydrodynamics   & GR & Newtonian & Newtonian & Newtonian \\
Neutrino opacities (see Table~\ref{tab:opac}) & Full & Full & Reduced & Reduced \\
Observer corrections & GR & $\mathcal{O}(v/c)$ & $\mathcal{O}(v/c)$ & None \\
Homologous core, \Mshock\ (\msun) & 0.429 & 0.492 &  0.717 & 0.427 \\
Central-core density at bounce, $\rho_{\rm c}$ ($ \times \, 10^{14}$~\gcc) & 4.714 & 4.264 &  3.336 & 3.157 \\
Central-core electron fraction ($Y_e$) at bounce & 0.2448 & 0.2407 &  0.3046 & 0.1855 \\
Central-core lepton fraction ($Y_L$) at bounce & 0.2804 & 0.2782 &  0.3696 & 0.2007 \\
Peak shock radius (km) & 162 & 190 & 171 & 142\\
Peak \nue-Luminosity (\Bethes) & 406 & 450 & 448 & 160
\enddata
\end{deluxetable*}

We present results from four spherically symmetric, core-collapse supernova models of decreasing physical fidelity.
The most physically complete model (\grfull, black lines in plots)  utilizes the more modern and complete \fullop\ opacities and the full general relativistic treatment of gravity, hydrodynamics, and transport as described in \citet{LiMeMe04}.
The first approximate model (\nfull, red lines) replaces the full general relativity of the \grfull\ model with Newtonian gravity, $\mathcal{O}(v/c)$ hydrodynamics, and $\mathcal{O}(v/c)$ transport.
The second approximate model (\nreduc, green lines) further replaces the more complete  \fullop\ neutrino opacity set with the \reducop\ opacities (see Table~\ref{tab:opac} for a full comparison), while retaining the Newtonian gravity, $\mathcal{O}(v/c)$ hydrodynamics, and $\mathcal{O}(v/c)$ transport.
This approximation includes  the important effect of removing neutrino weak interactions that down-scatter the neutrino energy.
The final approximate model (\noc, blue lines) retains the Newtonian gravity and $\mathcal{O}(v/c)$  hydrodynamics of the previous model, but drops the observer corrections completely, reducing the transport to $\mathcal{O}(1)$.
[N.B. The $\mathcal{O}(1)$ and $\mathcal{O}(v/c)$ hydrodynamics equations are identical.]
The general and core-bounce properties of all models are summarized in Table~\ref{tab:models} and plotted in Figure~\ref{fig:bounce}.

\subsection{GR versus Newtonian Gravity \label{sec:GR}}

\begin{figure}
\epsscale{1}
\plotone{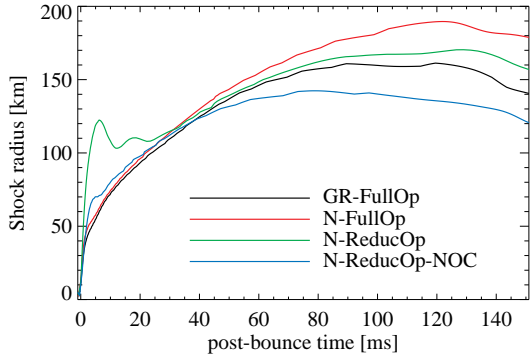}
\caption{Shock trajectories in km, versus time after bounce, for all models. The colors have the same meaning as in Figure~\ref{fig:bounce}. Shock position is computed by bisecting the pair of mass shells with the largest negative radial velocity gradient $-\partial v_r/\partial r$.  \label{fig:shock}}
\end{figure}

\begin{figure}
\epsscale{1}
\plotone{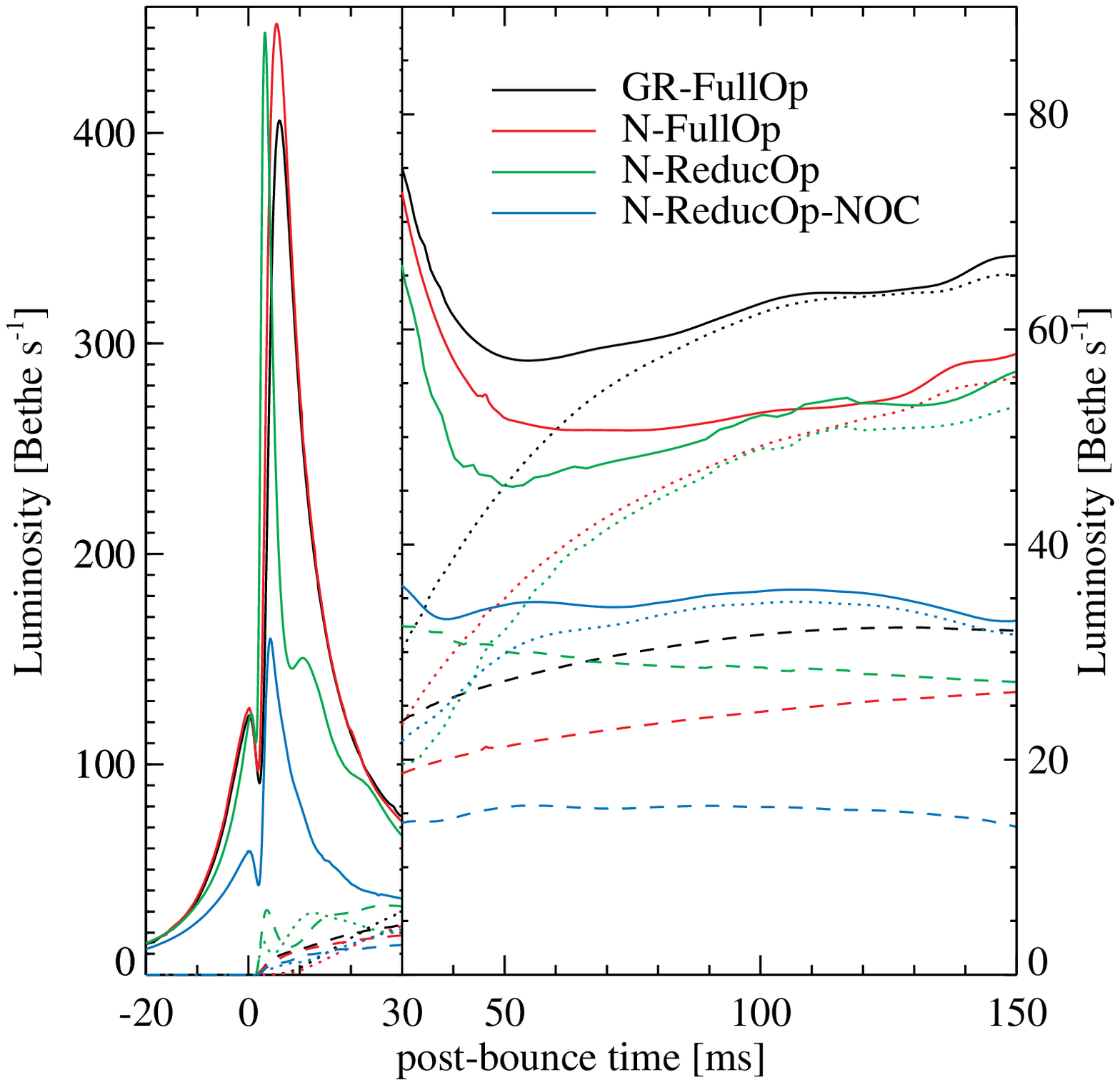}
\plotone{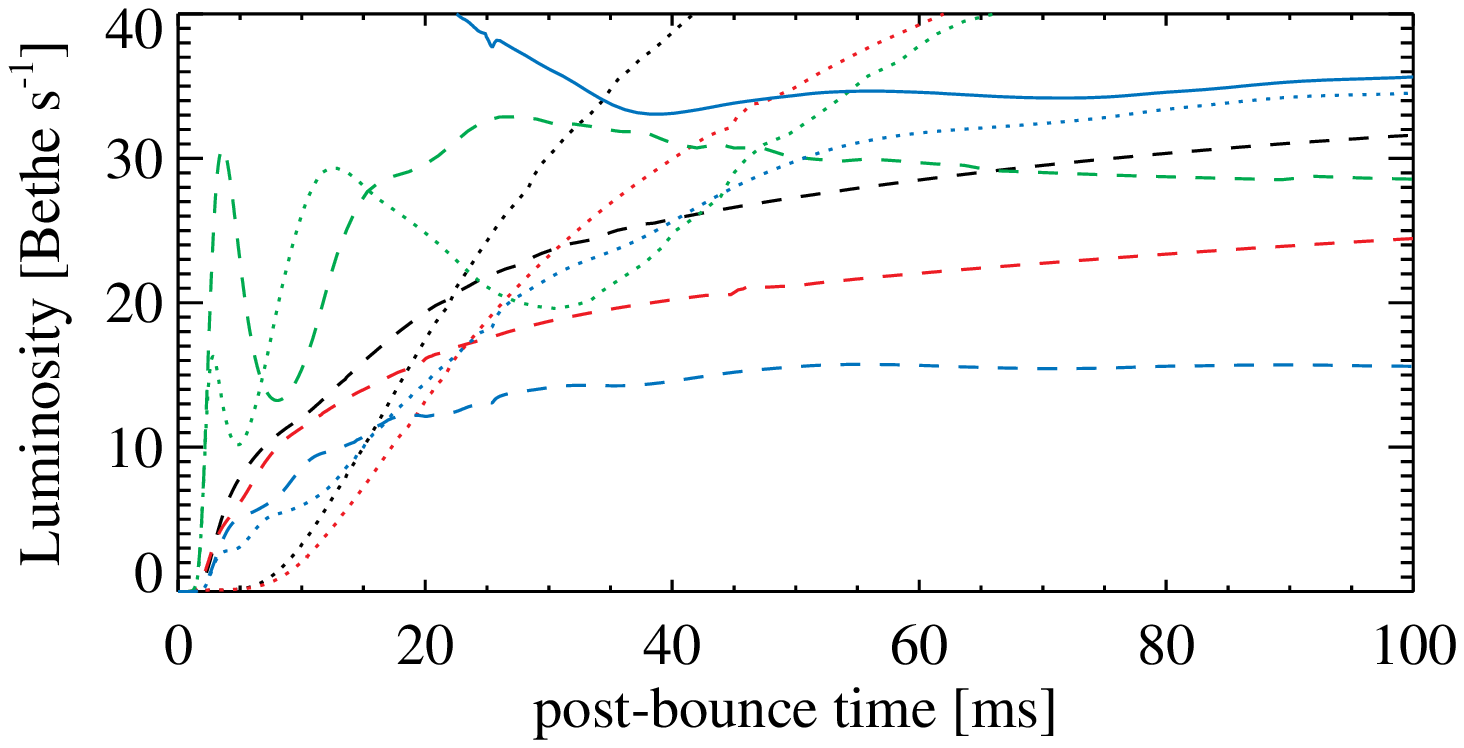}
\caption{Comoving-frame neutrino luminosities measured at 400~km for all models. Colors are as in Figure~\ref{fig:bounce}.
 Electron neutrino, \nue, luminosities are represented by solid lines, \nuebar-luminosities by dotted lines, and \numt-luminosities by dashed lines.
\numtbar-luminosities are indistinguishable from \numt-luminosities, and omitted from this figure.
The luminosities are in \Bethes, where 1~Bethe = $10^{51}$~ergs.
The lower panel provides a detailed view of the luminosities below 40~\Bethes\ during the first 100~ms after bounce. \label{fig:lumin}}
\end{figure}

\begin{figure}
\epsscale{1}
\plotone{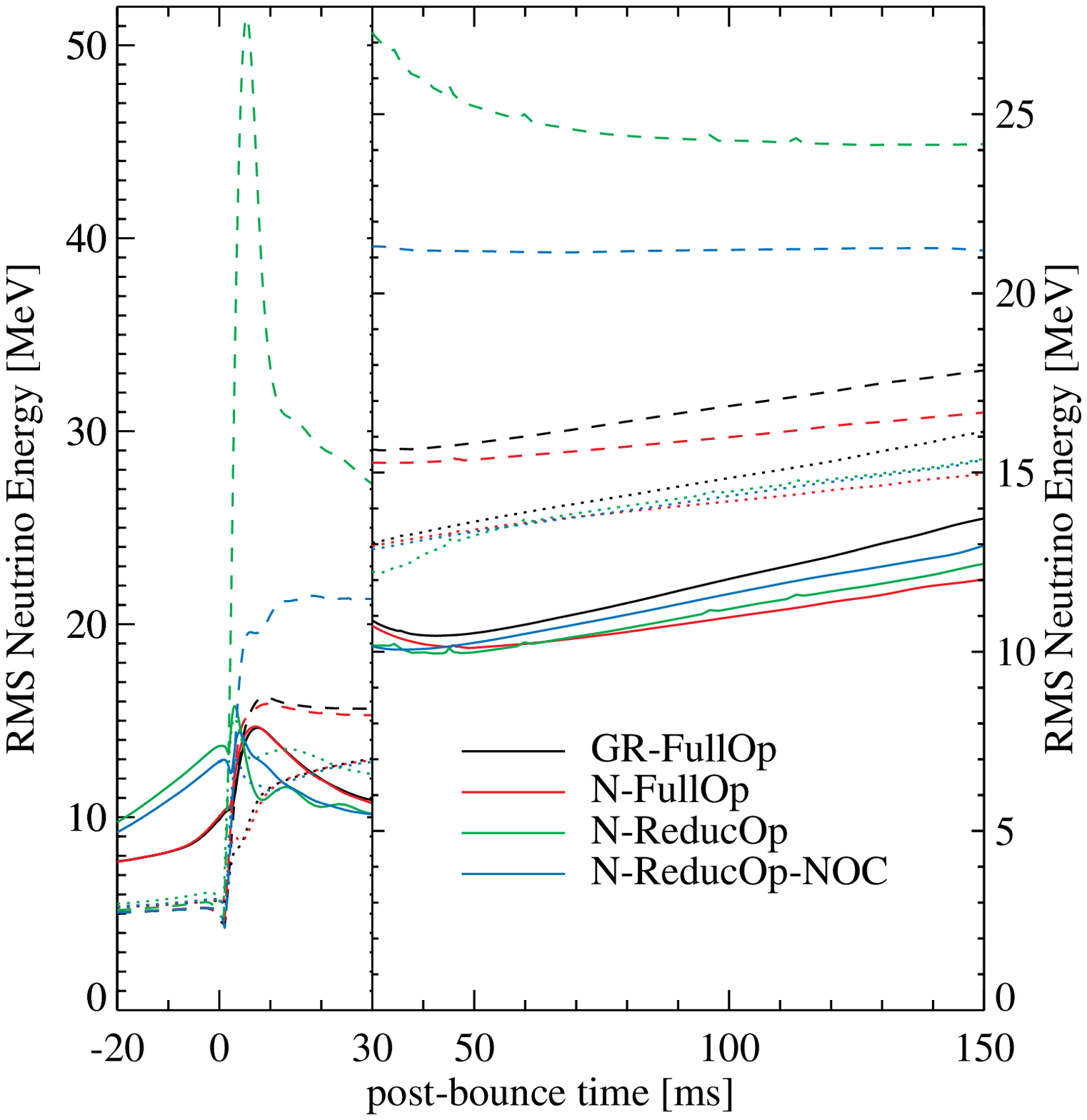}
\plotone{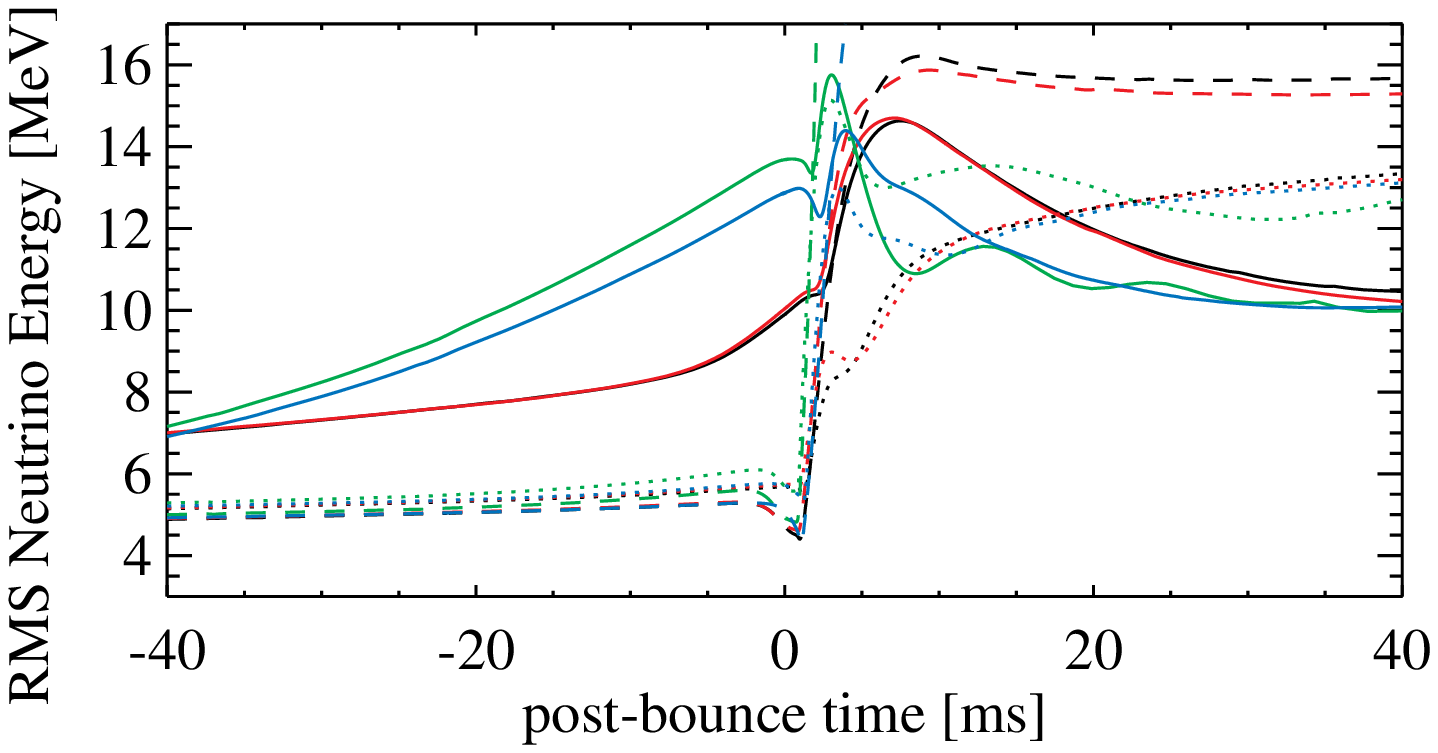}
\caption{Comoving-frame neutrino RMS energies, $\meanE{\nu} = (\int d\mu \, dE \, E^4 F/ \int d\mu \, dE \, E^2 F)^{1/2}$, measured at 400~km for all models. 
RMS energy is computed over number density, not number flux.
Colors are as in Figure~\ref{fig:bounce}.
Line styles are as in Figure~\ref{fig:lumin}.
The lower panel provides a detailed view of $\meanE{\nu}$ for values less than 20~MeV over the perious $\pm40$~ms. \label{fig:avgen}}
\end{figure}

The effects of general relativity on the core dynamics are seen in the comparison of the first two models (\grfull\ and \nfull).
The deeper gravitational well of the GR model results in a more compact homologous core at bounce (0.429~\msun\ versus 0.492~\msun) with a higher central density ($4.71 \times 10^{14}\,\gcc$ versus $4.26 \times 10^{14}\, \gcc$) and higher temperatures throughout the unshocked core (Figure~\ref{fig:bounce}).
The electron ($Y_e$) and  lepton ($Y_L$) fractions are essentially unchanged modulo the shift in shock position, as are the velocity and entropy, while the pressure differences follow the density differences.
As the shock moves out, the shock radius for both models (Figure~\ref{fig:shock}) remains close for the first 40~ms after bounce and then diverges.
The \grfull\ model  has maximum shock extent that is 30~km (20\%) smaller than the \nfull\ model, and by 150~ms after bounce the shock radius is 40~km (30\%) smaller.
Several quantities reflect the long-term effect of the more compact, and therefore hotter, proto-NS in the \grfull\ model, including the  higher luminosities for all neutrino species (Figure~\ref{fig:lumin}) and the higher RMS energies (\meanE{\nu}) of neutrinos of all flavors after the break-out burst (Figure~\ref{fig:avgen}).
These differences are in accord with those already reported by \citet{LiMeTh01}, \citet{BrDeMe01}, and \citet{BuRaJa06} using  different progenitors,  different opacity sets (similar to \reducop, though including NES),  different energy and angle resolutions, and for the latter two cases, different codes.
Our GR/Newtonian comparison is included here for completeness and to facilitate relative comparisons across all four models.

\subsection{Reduced Neutrino Opacities\label{sec:reducop}}

The changes induced as we go  from the \fullop\ opacities  (model \nfull)  to the \reducop\ opacities (model \nreduc) in the Newtonian-gravity,  $\mathcal{O}(v/c)$-hydrodynamics, and  $\mathcal{O}(v/c)$-transport limit are more dramatic than those seen for the transition from models \grfull\ to \nfull\ in \S\ref{sec:GR}.
The shock position at bounce changes from 0.492~\msun\ for \nfull\ to 0.717~\msun\ for \nreduc\ (Figure~\ref{fig:bounce}), with the entropy peak (upper right) making the same shift.
The increase in the initial shock mass, \Mshock, is correlated with the corresponding increase in core lepton fraction, from $Y_L = 0.28$ to 0.37 ($\Mshock \propto Y_L^2$).
The larger \Mshock\ for \nreduc, relative to the other models, results in a correspondingly larger region of high pressure, temperature, and density at bounce.
The  vigorous post-bounce shock of model \nreduc\  results in a  strong ``ringing'' of the shock (Figure~\ref{fig:shock}).
\citet{ThBuPi03} reported a similar ringing for their ``no NES'' model.

The \nue-luminosity of the \nreduc\ model reaches the same peak value as in the \nfull\ model, 450~\Bethes, but the breakout burst is much shorter in duration and represents a smaller total emission of \nue.
The shock starts out at a larger mass coordinate and passes through less total mass before becoming a steady accretion shock.
Like \citet{ThBuPi03}, we see oscillations of the $\nu$-luminosities and \meanE{\nu}\ (Figures~\ref{fig:lumin} \& \ref{fig:avgen}) just after bounce induced by shock oscillations passing through the neutrinospheres.

The differences between the \nfull\ and \nreduc\ models can be understood by considering three  opacity changes {\it imposed simultaneously}: (A) the inclusion of NES/NPS; (B) the use of the LMSH EC table; and (C) the use of the \citet{RePrLa98} nucleon opacities.

(A) The effects of omitting the NES opacity {\it alone} during collapse were explored by \citet{MeBr93c}, who showed that energy down-scattering by NES allowed the energy down-scattered neutrinos to escape more easily  because of the lower absorption and scattering cross-sections at lower energies, and reduced the core $Y_e$ by 15\% and neutrino fraction, \Ynu, by 30\% in their model with NES relative to one without NES.
The higher number of trapped neutrinos, without NES, is reflected in the higher core $Y_e$, $Y_L$, and \Ynu\ for our model \nreduc.
The ``no-NES'' model of \citet{ThBuPi03} (\S7.4, Figures 20 \&~21) also shows  large differences in \meanE{\numt}, with a bounce ``spike'' reaching 32~\mev, and a \meanE{\numt}\ increase at 150~ms post-bounce of 4~\mev\ (20\%) relative to a model with NES.
This compares to a 50~\mev\ ``spike'' and 7~\mev\ (40\%) increase at 150~ms after bounce in \meanE{\numt}\ for our \nreduc\ model   relative to our \nfull\ model.

(B) We have removed the LMSH EC table from \reducop\ opacities in favor of the simpler IPA prescription, as not all modern supernova simulations use EC rates like the LMSH EC table, which reflect the ensemble of nuclei and their excited states in the collapsing core.
\citet{HiMeMe03} found that the enhanced EC arising from the removal of the artificial cut-off in IPA for heavier nuclei that occur at higher densities during collapse  decreased the central-core $Y_e$  and \Mshock\  at bounce by 10\% and 20\%, respectively, relative to the IPA implementation.
Conversely, IPA overestimates EC where it is active and leads to  excess  deleptonization and stronger collapse in the outer regions of the Fe-core where the $N \geq 40$ cut-off criterion is not triggered.
One such region can be seen at bounce outside the homologous core near 0.9~\msun\  (Figure~\ref{fig:bounce}), where lower $Y_e$ and higher density exist in the \nreduc\ model relative to the \grfull\ and \nfull\ models.

(C) We have also replaced the $\nu$-nucleon emission, absorption, and scattering opacities \citep{RePrLa98} in \fullop\ with the corresponding opacities of \citet{Brue85} to eliminate neutrino-energy down-scattering on nucleons.
We have previously found \citep{LeHiBa10} that the inclusion of the enhanced nucleon opacities results in an enhancement of the luminosities, but not RMS energies, and lifts the post-bounce shock outward by 10~km by 100~ms post-bounce through absorption of the excess luminosity.
These findings are consistent with the results of \citet{RaBuJa02} on the enhanced neutrino--nucleon opacities.

\subsection{No Observer Corrections}

\begin{figure*}
%\epsscale{0.95}
\plotone{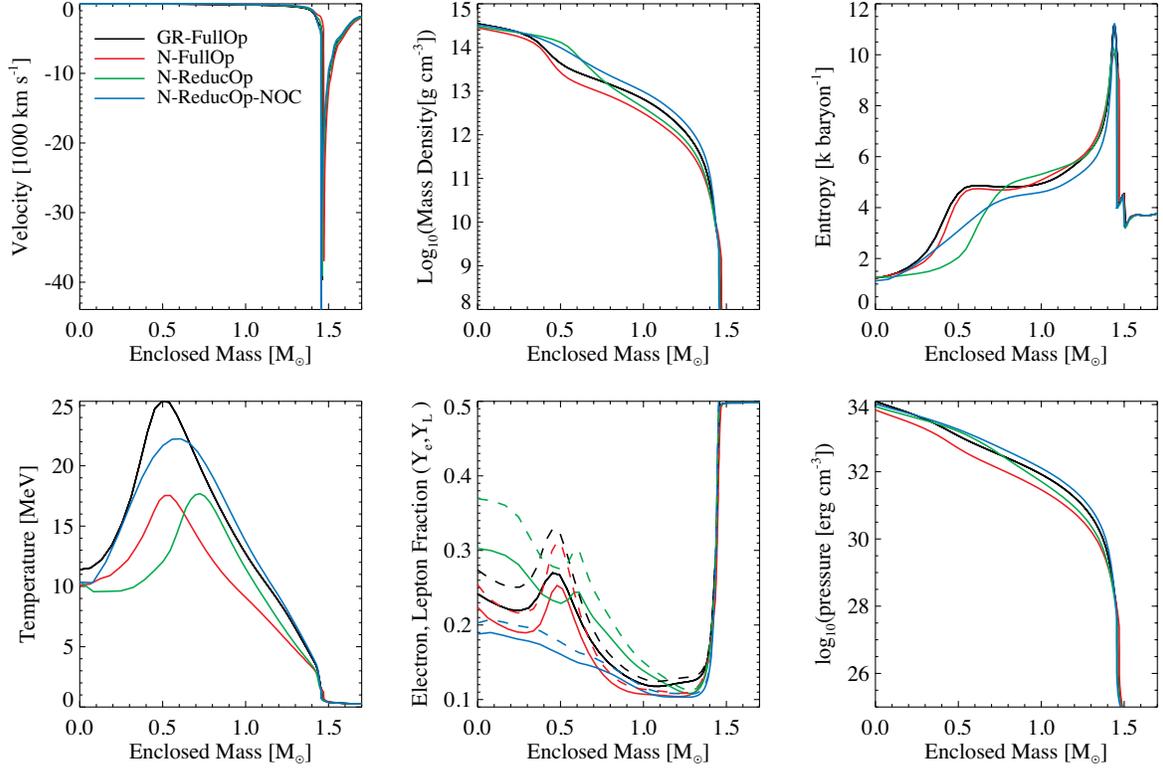}
\caption{Same as in Figure~\ref{fig:bounce}, but at 100~ms after core bounce. \label{fig:100mass}}
\end{figure*}

\begin{figure*}
%\epsscale{0.95}
\plotone{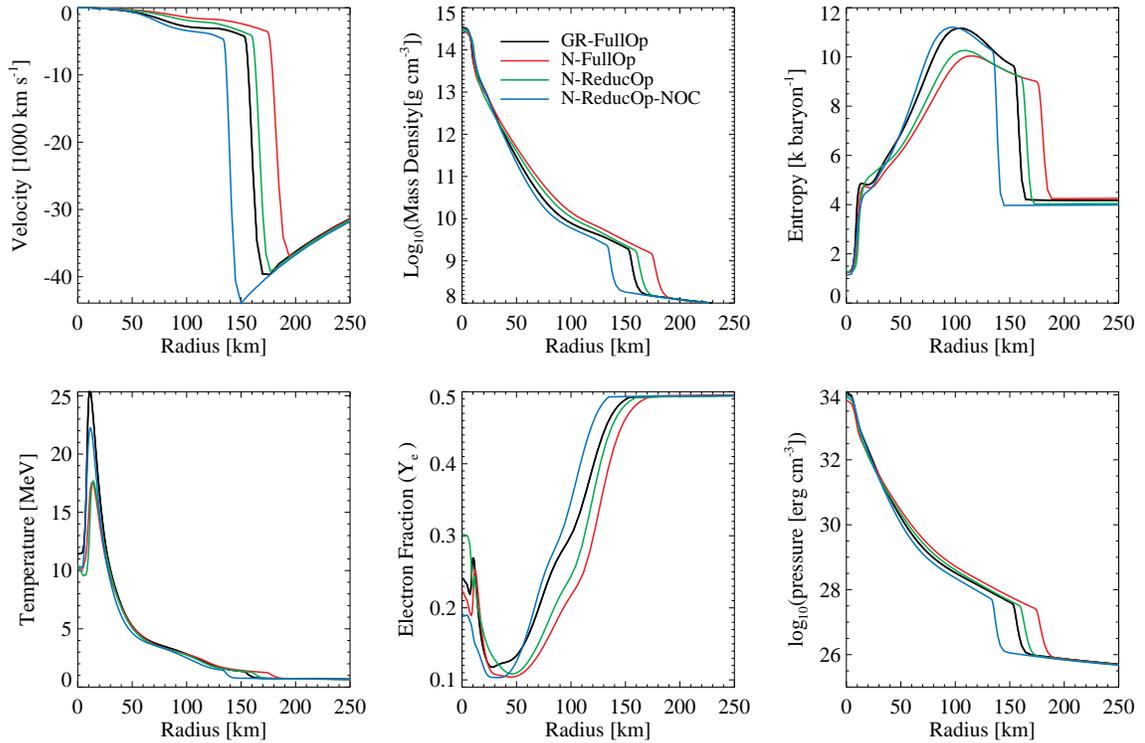}
\caption{Same as in Figure~\ref{fig:100mass}, but as a function of radial coordinate, $r$, in km. Net lepton number has been omitted, as $Y_L \approx Y_e$ for all but the inner few km. \label{fig:100ms}}
\end{figure*}

For the final comparison we change the treatment of the observer corrections in the transport equation.
In model \noc\ we have removed the velocity-dependent observer corrections from the Boltzmann transport equation in the appropriate, Lagrangian approach as described by \S\ref{sec:noc}, equation~(\ref{eq:lagnoc}), but retain the Newtonian gravity, $\mathcal{O}(v/c)$-hydrodynamics, and reduced opacities of the \nreduc\ model.

Dropping the observer corrections in model \noc\ results in a dramatic change in the properties of the core at bounce, as can be seen in Figure~\ref{fig:bounce} and Table~\ref{tab:models}.
The homologous core mass drops from 0.717~\msun\ for the \nreduc\ model to 0.427~\msun\ for the \noc\ model; the latter of which is virtually indistinguishable from the homologous core mass, \Mshock, for the most physically complete model (0.429~\msun\ for \grfull).
This coincidental alignment of the bounce shock positions for the most and least physically complete models should be contrasted with the lower electron and lepton fractions ($Y_e, Y_L$) and density (see Table~\ref{tab:models}) in the homologous core for the \noc\ model relative to the \grfull\ model, as well as the larger inflow velocities and densities and lower $Y_e$ and $Y_L$ outside the shock, which imply, among other things, an increased ram pressure against  which the shock must propagate.

The  shock (Figure~\ref{fig:shock}) in the \noc\ model starts  more vigorously  than in the  \fullop\ models, but does not show the large overshoot of the other reduced opacity model, \nreduc.
All of the shock trajectories cross near 35~ms after bounce, with the \noc\ model having the deepest shock throughout the rest of the run.
This is in stark contrast to the other Newtonian models, which have larger shock radii relative to the \grfull\ model.

The neutrino luminosities (Figure~\ref{fig:lumin}) of the \noc\ model are also substantially lower than for any other model.
The \nue-luminosity from shock-breakout peaks at 160~\Bethes\ for the \noc\ model relative to the 400--450~\Bethes\ for the  models with observer corrections.
The \nuebar-luminosities of all models approaches the \nue-luminosities at around 80~ms after bounce, and track together thereafter.
By 150~ms after bounce, the \nue\nuebar-luminosities for the \noc\ model are approximately 32~\Bethes, while the other two Newtonian models have \nue\nuebar-luminosities of $\sim 55\,\Bethes$ and the \grfull\ model has \nue\nuebar-luminosities of $\sim 60\,\Bethes$.

The RMS neutrino energies (Figure~\ref{fig:avgen}) for \nue\ (solid lines) and \nuebar\ (dotted lines) in the \noc\ model follow those in the \nreduc\ model closely, with \meanE{\nue}\ slightly higher before bounce and at later times, except in the immediate post-bounce period when it oscillates in the \nreduc\ model with the shock.
In contrast, the post-bounce \meanE{\numt}\ for model \noc\ is  substantially lower relative to model  \nreduc, and the ``spike'' after bounce is gone.
After breakout both \reducop\ models have large \meanE{\numt}\ relative to the \fullop-models.

As we traverse our four models, it is clear that all of the neutrino luminosities are significantly affected.
The general trend is for the luminosities to decrease considerably as we go from \grfull\ to \nfull\ to \nreduc\ to \noc.
The largest variations among the models are exhibited by the electron-flavor neutrinos, with luminosity variations as large as 35~\Bethes\ at 150~ms after bounce.
However, the variations in the \meanE{\nue}\ and \meanE{\nuebar}\ are not as dramatic as we traverse the four models, and not monotonically decreasing as the model sophistication decreases.
Variations in the \meanE{\numt}, like their luminosity counterparts, remain significant, although not monotonically decreasing with model.
The post-bounce RMS energies, \meanE{\numt}, vary by  $\gtrsim10\,\mev$ up to 150~ms after bounce.

At 100~ms after bounce (Figure~\ref{fig:100mass}) the shock encloses 1.45~\msun\ for all four models.
The most significant differences seen in the dense core are in $Y_e$ and $Y_L$ (lower center), where the \noc\ model has generally the lowest $Y_e$ and $Y_L$ and lacks the peak the other models exhibit just outside their bounce shock positions, \Mshock.
Figure~\ref{fig:100ms} shows the same data, but focuses on the outer, ``hot-mantle'' region between the shock and  proto-NS with the most evident differences being those related to the shock radius (smallest for \noc). 
Though shifted in radius, the hot-mantle region is similar among all models.

\begin{figure}
\epsscale{1}
\plotone{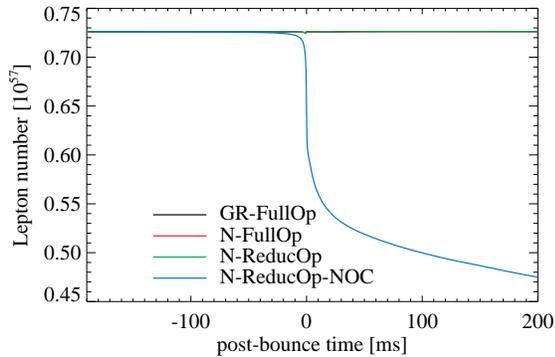}
\caption{Lepton number on the grid plus neutrino flux through the outer boundary, \lepcons, indicating the quality of numerical lepton conservation for all models. Model \noc\ loses 34.7\% of the original \lepcons\ by 200~ms after bounce, while the other three models conserve \lepcons\ to within 0.1\%. \label{fig:leptoncons}}
\end{figure}
The \noc\ model shows (Figure~\ref{fig:leptoncons}) a drop in total conserved lepton number  (\lepcons, the lepton number on the computational grid plus the time-integrated number flux of neutrinos at the outer boundary) starting just before bounce and continuing throughout the rest of the run.
This is not seen in the other models, which maintain lepton conservation.
The root of the non-conservation can be seen in the integration of equation~(\ref{eq:lagnoc})  for neutrino number over the entire grid.
It results from the $(F/\rho) (\partial \rho/\partial t)$ ``compression'' term and is strongest during the epoch of high $Y_\nu$ and  rapid density changes surrounding core bounce.

\begin{figure}
\plotone{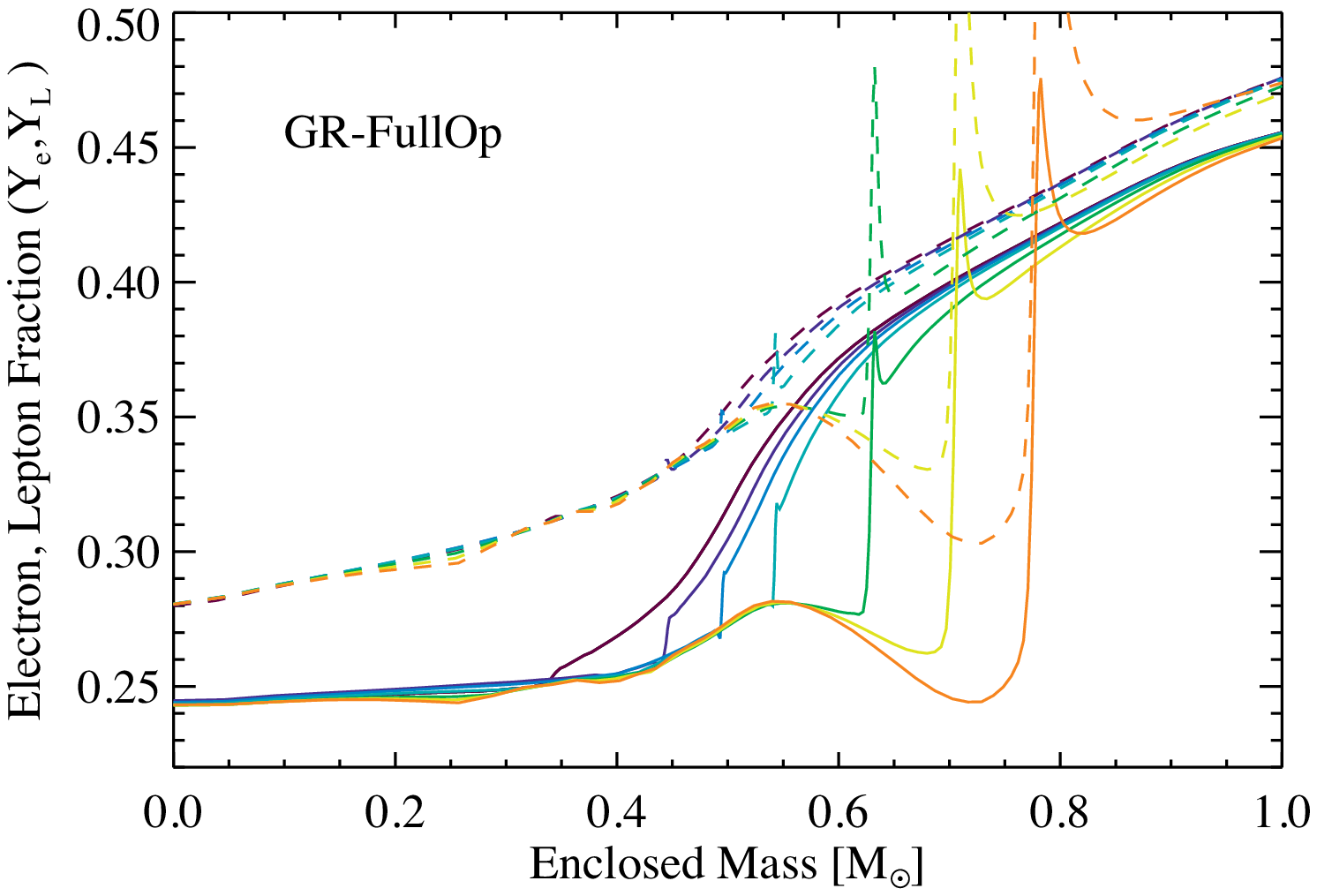}
\plotone{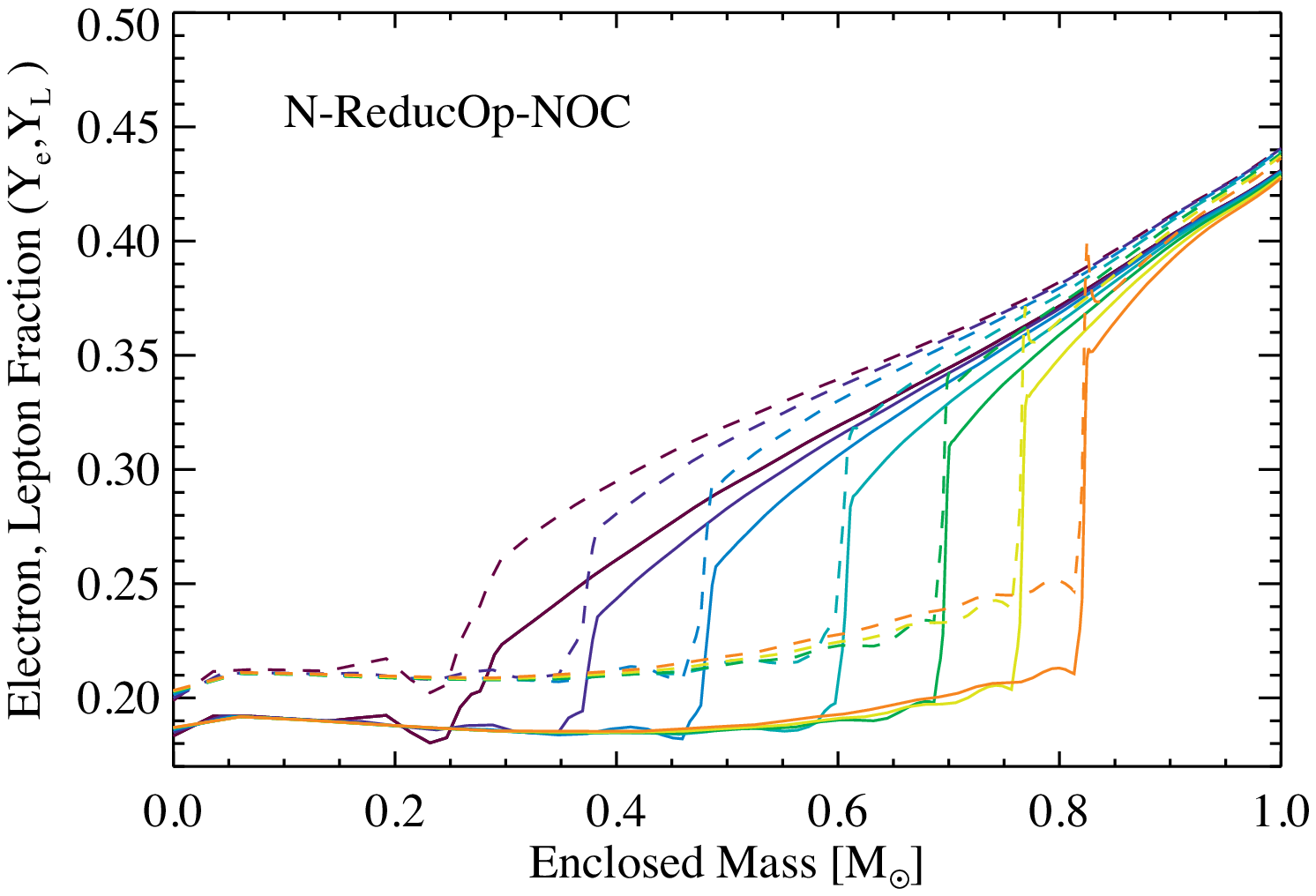}
\caption{
Sequence of electron (solid lines) and lepton (dashed lines) fraction profiles for the \grfull\ (upper panel) and \noc\ (lower panel) models showing the formation of the shock and the shock progress through bounce into the breakout phase.
The colors represent different epochs, which are equally spaced in computational time step with time advancing from dark to light gray (violet to red in the electronic edition). \label{fig:breakout}
}
\end{figure}

We  illustrate this loss through $Y_e$ (Figure~\ref{fig:breakout}, solid lines) and $Y_L$ (dashed lines) profiles for the \grfull\ model (upper panel) and the \noc\ model (lower panel) as a series of temporal snapshots near bounce.
Until the shock reaches 0.5~\msun\ in the \grfull\ model (upper panel), the core is still opaque, and the neutrinos are trapped.
Therefore, total lepton number is conserved locally, and $Y_L$ is steady inside 0.5~\msun\ during shock breakout as depicted in the upper panel of Figure~\ref{fig:breakout}.
When the shock reaches 0.5~\msun, it begins to break through the neutrinospheres, and the neutrinos can escape, causing the local $Y_L$ and $Y_e$ to drop behind the shock.
The escaping neutrinos contribute to $Y_L$ in front of the shock as a visible pulse, a small portion of which are absorbed by the cold, infalling matter ahead of the shock, forming a transient radiative precursor in $Y_e$.
For the \noc\ model (lower panel) there is no such corresponding epoch of local lepton conservation as the shock forms in the opaque (neutrino trapped) inner core before emerging through the neutrinospheres.
The net effect of the compression term in equation~(\ref{eq:lagnoc}) is one of destroying neutrinos, which then results in a net decrease in $Y_e$ via the interactions $e^- p \leftrightarrow \nue n $ and $e^+ n \leftrightarrow \nuebar p$.
The loss of neutrinos to the compression term reduces the neutrino pulse ahead of the shock and lowers the \nue-luminosity in the breakout burst  (Figure~\ref{fig:lumin}).

The fundamentally different behavior of the \noc\ model  stems from two factors: 
(1) the omission of the energy derivative term in equation~(\ref{eq:boltzlag}) or the equivalent term in equation~(\ref{eq:boltzeulov/ccon}); and 
(2) the fact that equation~(\ref{eq:lagnoc}) is manifestly non-conservative for neutrino, and consequently lepton, number when integrated over mass.
In the neutrino opaque regions, the energy-derivative term is responsible for promoting neutrinos in energy as they are compressed, as expected for a relativistic Fermi gas and first noted by \citet{Cast72} and \citet{Arne77}.

\section{Contemporary Multidimensional Supernova Modeling\label{sec:codes}}

\subsection{Multidimensional Supernova Codes}

There are five extant codes that can compute the spectral neutrino radiation hydrodynamics for core-collapse supernova simulations in 2D or 3D.
These codes are (in alphabetical order by name): the FAU-NCSU-Oak Ridge 2D/3D code \chimera\ (S.~W. Bruenn et al., in preparation), the Stony Brook  2D code \vtd\ \citep{SwMy09,SwMy05}, the MPA-Garching  2D code \vertex\ \citep{RaJa02,BuRaJa06}, the Arizona-Caltech-Hebrew University-Princeton 2D code \vulcan\ \citep{LiBuWa04,BuLiDe07}, and the Kyoto-Tokyo 2D/3D  \zidsa\ code \citep{SuKoTa10}.

Of the multi-D codes, \chimera\ and (P{\sc rometheus}-)\vertex\ include a spherically-symmetric, post-Newtonian GR approximation, while the others are strictly Newtonian in their gravitation, hydrodynamics, and neutrino transport.
\citet{MuJaDi10} have updated ({\sc CoCoNuT}-)\vertex\ to include general relativity in the transport and hydrodynamics using the conformally flat approximation.

\chimera, \vtd, and \vulcan\ transport neutrinos by the flux-limited diffusion method (FLD).
\vulcan\ also has a non-moment, multi-angle ($S_N$) mode.
\vertex\ uses the variable Eddington tensor (VET) method with a closure computed using a spherically averaged, model Boltzmann equation.
The \zidsa\ code uses the Isotropic Diffusion Source Approximation  \citep[IDSA;][]{LiWhFi09}, which divides the neutrinos into ``trapped' and ``free-streaming'' neutrinos, with a diffusion source to connect them.

Of these codes, only \vtd\ is capable of solving the full space-neutrino energy-species coupling of the neutrino transport  that the core-collapse supernova problem requires, while all other codes break at least one aspect of that coupling to reduce computational costs and simplify code development.
\chimera, \vertex, and \zidsa\ break the non-radial (lateral, or angular) spatial coupling through the ``ray-by-ray'' (RbR) approximation, and \vulcan\ breaks the coupling between energy groups and neutrino species.

\begin{figure}
\epsscale{.3}
\plotone{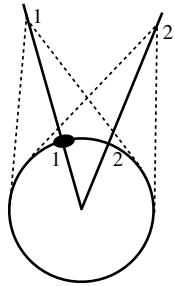}
\caption{Illustration of the ``ray-by-ray'' transport approximation. The circle represents the neutrinosphere and the solid lines represent two independent ``rays''  in the RbR approximation. The dashed lines are tangents to the neutrinosphere and indicate the regions that contribute to the neutrino field at points 1 and~2.
The ``blob'' on the neutrinosphere below point~1 is a ``hot spot'' where the temperature is higher than  the rest of the neutrinosphere.
For point~1, the RbR method will compute the neutrino field  as if the entire neutrinosphere has the properties of the hot spot, overestimating the neutrino flux and heating.
For point~2, the RbR misses the contribution of the hot spot by assuming the neutrinosphere properties are only those of the cooler region directly below it, underestimating the neutrino flux and heating.
 \label{fig:rbr}}
\end{figure}

In the RbR approximation,  the neutrino transport is computed as a number of independent, spherically symmetric problems, referred to as ``rays,'' which allows for the reuse of existing 1D neutrino transport codes.
(See Figure~\ref{fig:rbr} for a schematic illustration of the RbR approximation.)
RbR methods exhibit good parallel scaling for large numbers of independent radial rays, which can be evolved without communication while computing the neutrino transport.
Typically, in RbR codes, the neutrinos in opaque regions are advected laterally with the fluid motions and contribute to the pressure.
The independence of the rays artificially sharpens the lateral variation in the neutrino luminosity and heating above the proto-NS, which results in some regions of the hot mantle being overheated and others underheated.
The transport studies of \citet{OtBuDe08} using \vulcan\ in  multi-angle mode showed that full multi-D FLD underestimates the lateral variation in the neutrino radiation field, whereas  RbR codes are expected to overestimate the lateral variation.
\citet{BuRaJa06} concluded from analysis of their RbR models that the transient lateral variations in  neutrino flux and heating were not very likely to have dynamical consequences for the evolution of their models.
The impact of the RbR approximation on the simulation outcomes is not precisely known, and proper testing will have to wait until one of the RbR codes is upgraded to include full lateral transport, as  no extant code is currently capable of computing in RbR and non-RbR modes and there are significant differences between extant RbR and non-RbR codes in other respects.

The authors of \vulcan\ have chosen to break the energy and species coupling rather than the lateral spatial coupling.
\vulcan\ implements computational parallelism by solving for 2D-spatially-coupled neutrino transport  for each  energy-species group independently, with  communication only after transport to integrate neutrino heating/cooling from all energy groups.
The consequence of this design choice is that \vulcan\  cannot easily include either NIS-driven coupling of energy groups or the coupling of energy groups through observer corrections,  nor can it utilize more parallel processing elements than it has energy--species groups.

\subsection{Opacity Approximations}

\chimera\ and \vertex\ include all of the  \fullop\ opacities plus  additional corrections for weak magnetism and ion-ion correlations.
\vertex\ also includes the neutrino-pair flavor-conversion process \citep{BuJaKe03}. 
\vtd\ uses the \citet{Brue85} opacities, which are similar to \reducop, but do include the energy down-scattering from  NES.
\vulcan\ omits all of the NIS scatterings in favor of their IS counterparts, as does the \zidsa\ code because energy-coupled scattering has not yet been developed for the IDSA transport method.
\vulcan, \vtd, and \zidsa\ use an IPA for EC on nuclei, which cuts off electron capture by nuclei when the mean neutron number  $N \geq 40$, and overestimates it above the cut-off, while \chimera\ and \vertex\ use the more accurate LMSH EC table.

Some multi-D supernova codes (\vertex, \vulcan) use a single species, $\nux = \{\numt,\numtbar\}$, to represent all of the heavy-lepton flavor neutrinos, while the \zidsa\ code omits them completely.
	
\subsection{Observer Corrections}

\chimera, \vtd, and \vertex\ include the  observer corrections in the neutrino transport.
In the \zidsa\ code, adiabatic compression is properly handled for the trapped neutrinos, and $\mathcal{O}(v/c)$ observer corrections are included for free-streaming neutrinos.
These codes use neutrino transport based on equation~(\ref{eq:boltzeul}), its equivalent to $\mathcal{O}(v/c)$, or its GR equivalent.
Only  \vulcan\ neglects the observer corrections entirely, by solving the neutrino transport based on equation~(\ref{eq:boltzeulnoc}).
[The transport equation quoted in \citet{LiBuWa04} also omits the $\mu_0 v \,\partial f/\partial t$-term, which is typically considered of $\mathcal{O}(v^2/c^2)$ and dropped from most $\mathcal{O}(v/c)$ transport solutions.]

\section{Conclusions}

We have examined the consequences of removing (1) GR effects, (2) non-isoenergetic scattering and detailed nuclear EC opacities, and (3)  observer corrections from spherically symmetric models of core-collapse supernovae. 
We have found that all of these changes, individually and especially when taken together, affect the progress of stellar collapse and the post-bounce evolution of the shock and core thermodynamic properties in significant ways, in constrast to the assessments made by \citet{BuLiDe06,BuLiDe07} and \citet{NoBuAl10}.
We have computed variations in the shock radius, neutrino luminosities, and neutrino RMS energies as large as 60~km, 35~\Bethes, and 10~\mev, respectively, across the four models considered here.

Omission of GR results in a less compact core and an unrealistically more favorable shock progression after bounce.
Eliminating non-isoenergetic scatterings and simplifying electron capture on nuclei drastically reduces the core deleptonization and expands the homologous core at bounce.
Omission of the observer corrections dramatically alters  core deleptionization, the shock position, and neutrino luminosities after bounce, in part resulting from a complete breakdown of lepton number conservation.

The lepton non-conservation and non-promotion of neutrino energy resulting from omitting observer corrections in our \noc\ model results in a compact, low-$Y_L$ core and a shock trajectory that is the least favorable of our models.
The artificial loss of lepton number, lower neutrino luminosities, and the consequent lower neutrino heating rate and shallower shock trajectory may explain the lack of neutrino-driven explosions in models computed with \vulcan\ \citep[see][]{BuLiDe07}, in contrast to the results reported by others \citep{MaJa09,BrMeHi09b,SuKoTa10,TaKoSu11}.

Moreover, the changes in $Y_{e}$ and $Y_{L}$, their gradients, and the entropy gradients that we see as we traverse the models shown here will change the location and strength of convectively unstable regions in the proto-NS and between the proto-NS and the shock. The lepton and entropy gradients in the proto-NS drive prompt convection, the entropy gradients between the proto-NS and the shock drive neutrino-driven convection, and these in turn seed and are seeded by the SASI. That is, the changes we have documented in this transport study have implications for all of the multidimensional phenomena we know to be important in multidimensional supernova models once spherical symmetry is broken.

All of the ingredients (1)--(3) above must be included in multidimensional simulations of core-collapse supernovae to ensure  physical fidelity.
Their omission is not the only approximation used in current multidimensional simulations, some of which (like the ray-by-ray approximation) are inadequately understood and need to be better understood or phased out.
Certainly, further examination of these approximations is required within the context of multidimensional simulations.

\acknowledgements

E.J.L. is supported by grants from the NASA Astrophysics Theory and Fundamental Physics Program (grant number NNH11AQ72I) and the NSF PetaApps Program (grant number OCI-0749242).
A.M. and W.R.H. are supported by the Department of Energy Office of Nuclear Physics; and A.M. and O.E.B.M. are supported by  the Department of Energy Office of Advanced Scientific Computing Research. M.L. is supported by the Swiss National Science Foundation (grant numbers PP00P2-124879 and 200020-122287).
This research was supported in part by the National Science Foundation through TeraGrid resources provided by National Institute for Computational Sciences under grant number TG-MCA08X010.
This research used resources of the Oak Ridge Leadership Computing Facility at the Oak Ridge National Laboratory, which is supported by the Office of Science of the U.S. Department of Energy under Contract No. DE-AC05-00OR22725.

\end{document}